\documentclass[aps,prr,superscriptaddress,twocolumn
]{revtex4-1}
\usepackage{graphicx}
\usepackage{amsmath}
\usepackage{mathtools}
\usepackage{amsfonts}
\usepackage{bm}
\usepackage[normalem]{ulem}
\usepackage[usenames, dvipsnames]{xcolor}
\usepackage[export]{adjustbox}

\usepackage{hyperref}
\hypersetup{colorlinks=true, linkcolor=blue, citecolor=blue, urlcolor=blue}

\begin{document}

\title{Nickelate superconductors: an ongoing dialog between theory and experiments}

\author{Antia S. Botana}
\affiliation{Department of Physics, Arizona State University, Tempe, AZ 85287, USA}

\author{Fabio Bernardini}
\affiliation{Dipartimento di Fisica, Universit\`a di Cagliari, IT-09042 Monserrato, Italy}

\author{Andr{\'e}s Cano}
\affiliation{
Univ. Grenoble Alpes, CNRS, Grenoble INP, Institut Néel, 25 Rue des Martyrs, 38042, Grenoble, France
}
\date{\today}

\begin{abstract}
After decades of fundamental research, unconventional superconductivity has recently been demonstrated in rare-earth infinite-layer nickelates. The current view depicts these systems as a new category of superconducting materials, as they appear to be correlated metals with distinct multiband features 
in their phase diagram. Here, we provide an overview of the state of the art in this rapidly evolving topic.   
\end{abstract}

\maketitle




\section{Introduction}

Unconventional superconductivity, understood as superconductivity beyond the electron-phonon paradigm, remains a defining problem in condensed matter \cite{
norman11}. 
The challenge is exemplified by the high-$T_c$ cuprates, whose superconducting and normal-state properties keep puzzling even 35 years after their discovery \cite{norman15}. 
To overcome this impasse, a ``reasoning-by-analogy" approach has been perceived as a promising strategy that can offer valuable insights and new perspectives.
As such, the discovery of iron-based superconductors reinvigorated the field \cite{hosono15,hosono20}, 
twisted bilayer graphene notably spanned the question beyond bulk systems \cite{jh2018,mcdonald19}, and nickelates joined the club last year after Hwang and collaborators reported superconductivity in Sr-doped NdNiO$_2$ thin films, the first for a nickel-oxide material \cite{hwang19a}. 

\begin{figure}[b!]
    \includegraphics[height=.16\textwidth]{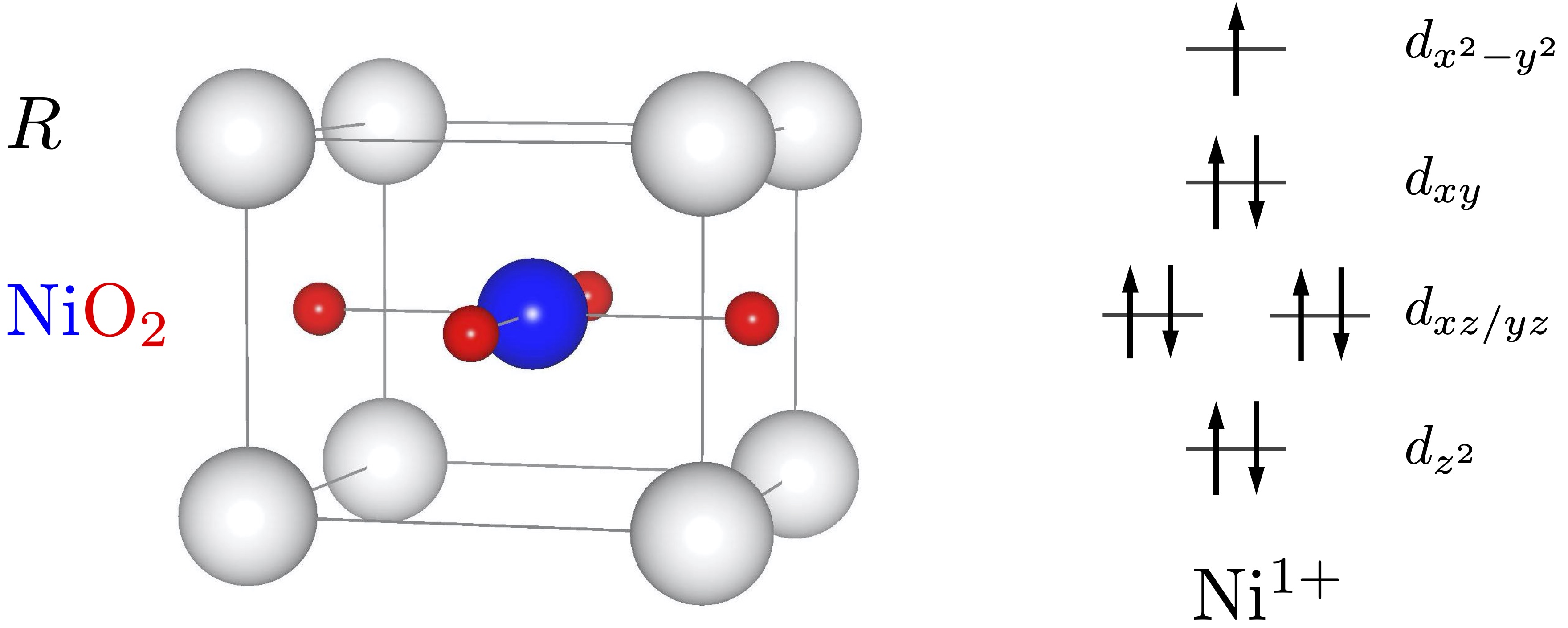}
    \caption{Ball-and-stick model of the unit cell of the infinite-layer nickelates $R$NiO$_2$ and sketch of the formal 3$d^9$ electronic configuration of a Ni$^{1+}$ atom in square-planar coordination. The set of $d$ orbitals is frequently divided into the subsets $e_g$-like~$= \{d_{x^2-y^2},d_{z^2}\}$ and $t_{2g}$-like~$=\{d_{xy},d_{xz},d_{yz}\}$ as in octahedral coordination. 
    }
    \label{fig:cartoon}
\end{figure}

Even though this latter discovery is very recent, nickelates have been debated as intriguing analogs to the high-$T_c$ cuprates for decades \cite{anisimov99,pickett-prb04,held09}. 
The analogy is most apparent precisely in  $R$NiO$_2$ ($R$ = rare-earth) layered materials (see Fig. \ref{fig:cartoon}). In these systems, the Ni atom features a nominal 1$^+$ oxidation state that formally provides the same 3$d^9$ electronic configuration of Cu$^{2+}$ in isostructural high-$T_c$ cuprates.
In reality, however, there are qualitative differences between the properties of these systems. 
LaNiO$_2$ for example is conducting ---even though weakly so--- with no magnetic order reported so far \cite{ikeda16, hayward99}, while the cuprates on the contrary are insulating antiferromagnets in their parent phases.
This clearly suggests that the embedding of the Ni atom in the actual crystal somehow spoils the tentative resemblance. 
Thus, the question is to what extent 
the analog $d^9$ picture for the nickelates survives the degree of covalency/metalicity of the actual atomic bonds, together with the eventual electronic correlations, and other important details of the overall electronic structure. 
It is the actual combination of these traits that defines the new physics that can eventually emerge in these nickel-based superconducting materials.

Here, we provide an overview on the current status of the field. 
Recent research has unveiled intriguing departures from the initial conjecture. 
Accordingly, superconducting nickelates are now depicted as a distinct class of correlated materials hosting unconventional superconductivity among other interesting properties. 

\begin{figure*}[t!]
    \flushleft{
    \hspace{27.6em} 
    {\scriptsize Nd$_{1-x}$Sr$_x$NiO$_2$}
    \hspace{9em}
    {\scriptsize Pr$_{1-x}$Sr$_x$NiO$_2$}
    }
    \\
    \centering 
    \includegraphics[height=.25\textwidth]{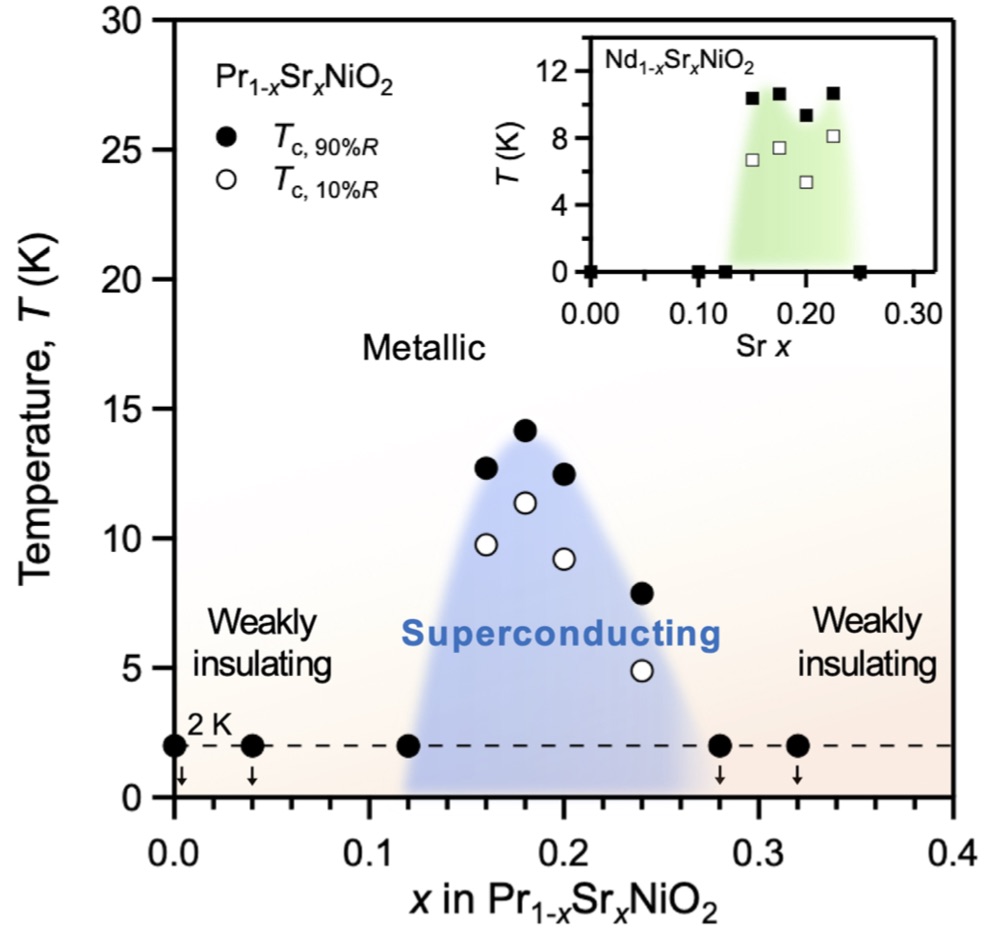}
    \hspace{4em}
    \includegraphics[height=.25\textwidth]{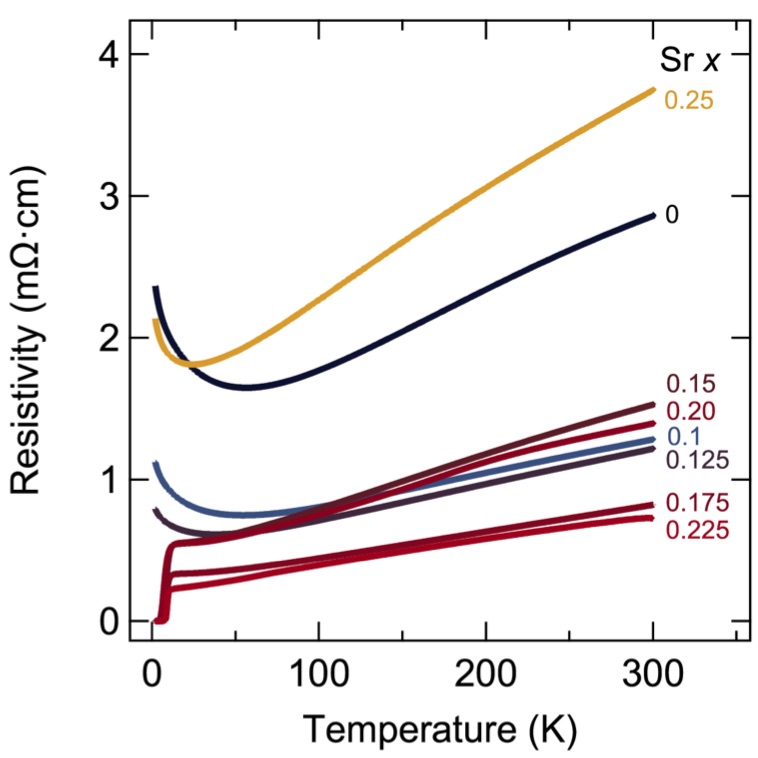}
    \hspace{2.5em}
    \includegraphics[height=.25\textwidth]{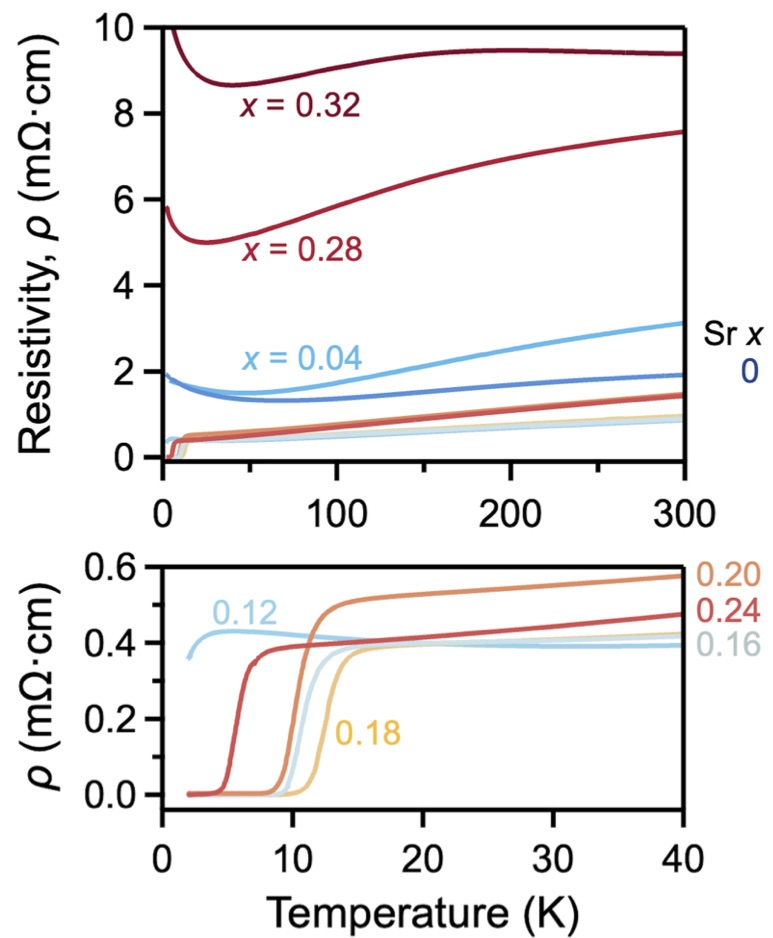}
    \caption{Temperature vs composition phase diagram of the superconducting infinite-layer nickelates reported so far ---{\it i.e.} Sr-doped NdNiO$_2$ and PrNiO$_2$ thin films on SrTiO$_3$ substrates--- and corresponding resistivity data (from \cite{hwang20-Ndphase-diagram} and \cite{hwang20Pr-b}). Sr-doping is equivalent to hole doping in these systems. The superconducting $T_c$ reaches 15 K in the best superconducting samples. The resistivity in the normal state shows metallic behavior as a function of temperature, with a Kondo-like upturn systematically observed for both underdoped and overdoped samples. The metallicity is however rather poor, and hence these systems can be seen as bad metals with weakly insulating features. At the same time, enhanced metallicity has been repeatedly reported in superconducting samples when compared to the non-superconducting ones.}
    \label{fig:phase-diagram}
\end{figure*}

\section{Experimental facts}

Rare-earth infinite-layer nickelates $R$NiO$_2$
have been known for decades. This special type of nicketales can be seen as the $n=\infty $ members of the series $R_{n+1}$Ni$_n$O$_{2n+2}$, 
with each member containing $n$-NiO$_2$ planes. LaNiO$_2$ was the first synthesized compound of this type in the early 1980s  \cite{crespin83}. However, it took 16 years to reproduce that synthesis, extend it to NdNiO$_2$, and perform a basic characterization of these unusual compounds \cite{hayward99,hayward2003,crespin05}.
This is a reflection of the challenging synthesis of these materials, typically achieved by first growing the nickelate in its $R$NiO$_3$ perovksite version, and then removing the apical oxygens from the NiO$_6$ octahedra with reducing agents such as hydrogen \cite{naito04}. This is the so-called topotactic reduction which, in practice, may have unwanted consequences such as hydrogen intercalation into the sample \cite{held20topotaticH}.  
In any case, no superconductivity has been found so far in this type of parent phases. This emerges only via charge carrier doping as in the cuprates, which represents an additional challenge in the case of the nickelates. In particular, hole doping the initial perovskite phase via Sr substitution on the $R$ site pushes the Ni valence from +3 towards an unstable +4. 
As such, doping and subsequent reduction has only been achieved by means of thin-film growth techniques (pulsed laser deposition (PLD) and molecular beam epitaxy (MBE)) \cite{hwang20aspects}, while superconductivity in doped bulk samples is yet to be reported \cite{li20absence,mitchell20bulk,wen20Sm}.

Fig. \ref{fig:phase-diagram} shows the temperature-composition phase diagram of (Nd,Sr)NiO$_2$ and (Pr,Sr)NiO$_2$ determined from resistivity data \cite{hwang20-Ndphase-diagram,hwang20Pr-b}. The measured resistivity as a function of temperature shows metallic behavior, with a low-temperature upturn systematically observed (Fig. \ref{fig:phase-diagram}). This upturn can be attributed to weak localization, or could be reminiscent of Kondo physics \cite{zhang20prbkondo}. 
At the same time, the room-temperature resistivity of these materials is comparatively high and it would result into insulating behavior in standard perovskite nickelates. 
In this sense, these systems can be seen as weak insulators or bad metals all along the phase diagram, in marked contrast to cuprates. 
Besides, no signature of long-range magnetic order has been reported so far for the parent infinite-layer nickelates \cite{ikeda16, hayward99, hayward2003}. 
However, a recent NMR study has pointed out the presence of antiferromagnetic fluctuations and quasi-static antiferromagnetic order below 40~K in Nd$_{0.85}$Sr$_{0.15}$NiO$_2$ \cite{yu20}.

The superconducting  $T_c$ upon hole doping reaches $\sim$ 15~K. However, the transport data reported so far is still sample dependent, with nominally equivalent samples displaying finite resistivity or a complete drop \cite{ariando20, hwang20-Ndphase-diagram, hwang20Pr-a}. Thus, there is hope for a higher $T_c$ in higher quality samples. Also, whether the original LaNiO$_2$ reference compound hosts superconductivity or not remains an important open question. As of now, superconductivity has only been reported in Sr-doped NdNiO$_2$ and PrNiO$_2$. This difference may be `simply' due to sample quality and/or the presence of topotactic hydrogen in LaNiO$_2$ ---{\it i.e.} the formation of LaNiO$_2$H instead of LaNiO$_2$--- as suggested from DFT calculations \cite{held20topotaticH}. Otherwise, it may be more intrinsically related to the rare-earth elements themselves ---{\it i.e.} closed vs open 4$f$ shells and the corresponding magnetic moments.

\begin{figure}[t!]
    \flushleft{
    {
    \hspace{4.6em}\scriptsize Nd$_{1-x}$Sr$_x$NiO$_2$}
    \hspace{6.6em}
    {\scriptsize Pr$_{1-x}$Sr$_x$NiO$_2$}
    }
    \includegraphics[height=.22\textwidth]{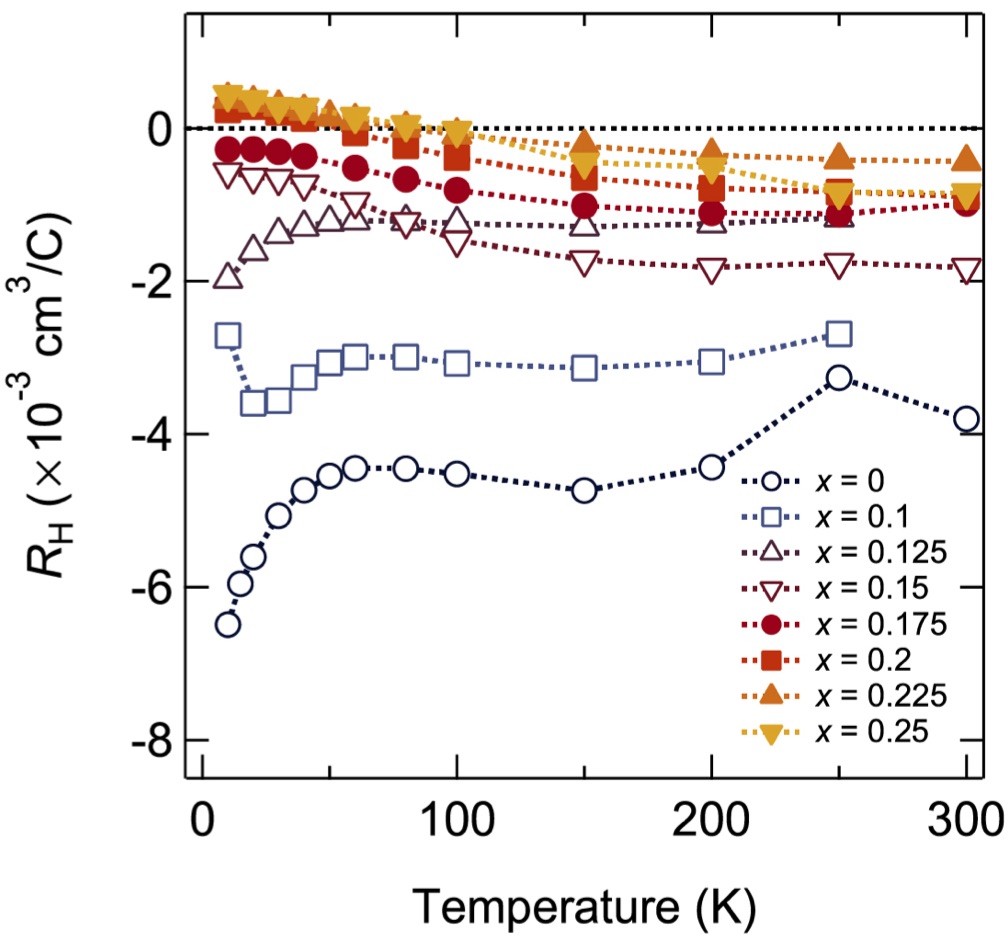}
    \includegraphics[height=.22\textwidth]{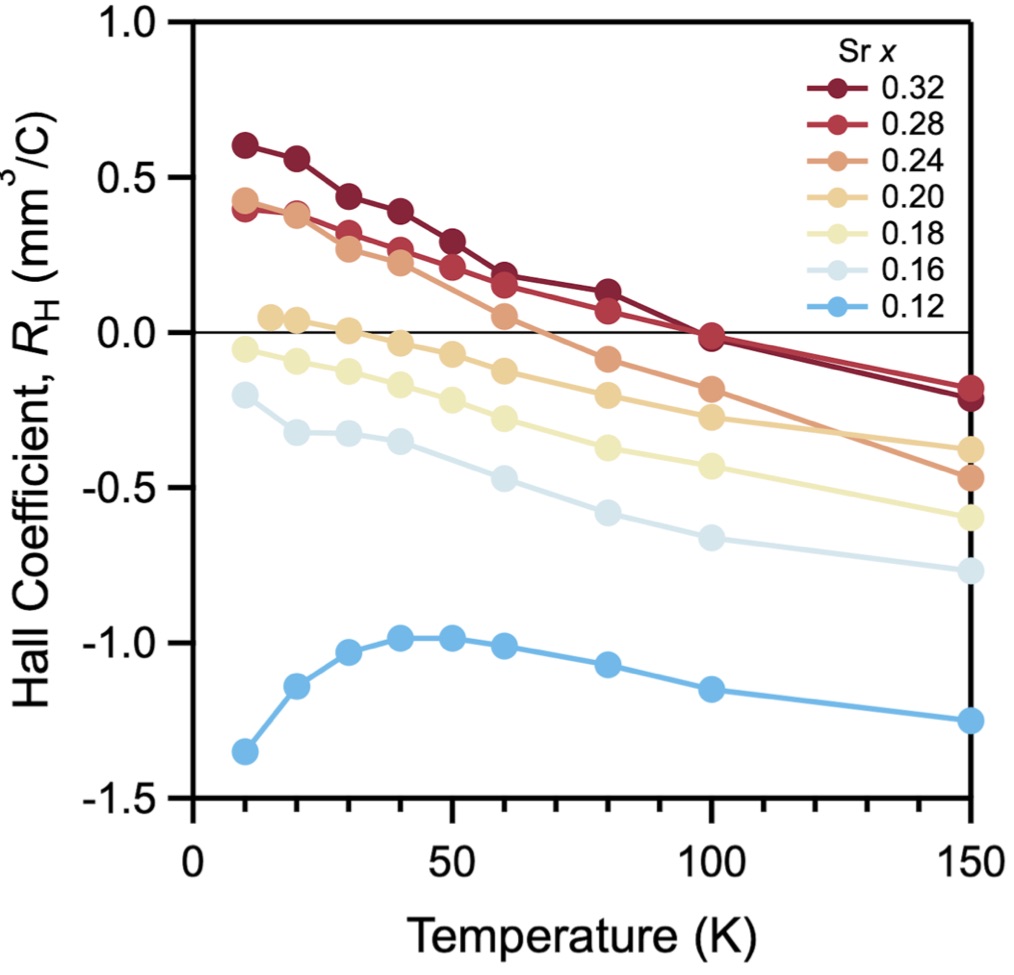}
    \caption{Hall coefficient measured in NdNiO$_2$ and PrNiO$_2$ as a function of temperature and Sr doping \cite{hwang20-Ndphase-diagram,hwang20Pr-b}.
    In both cases, the Hall coefficient below 100~K changes sign as a function of Sr content ($x$ in $R_{1-x}$Sr$_x$NiO$_2$). The main charge carriers have electron character in the parent compounds (x=0), which eventually changes to holes upon increasing doping. 
    This was the first experimental hint of an underlying multiband picture. }
    \label{fig:Hall}
\end{figure}

Fig. \ref{fig:Hall} illustrates the measured Hall coefficient for $R$NiO$_2$ that has been observed to change sign both as a function of temperature and as a function of doping. Accordingly, at low temperatures, the main charge carriers change from electrons in the parent compounds, to holes in the superconducting and over-doped samples. This evidences an underlying multiband character in infinite-layer nickelates, which is another important fundamental difference when compared to cuprates.  

\begin{figure}[t!]
    \includegraphics[height=.2\textwidth]{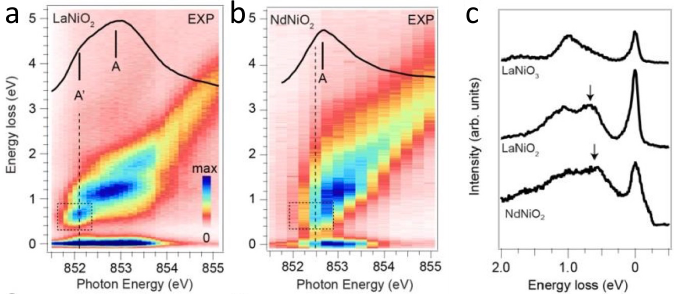}
    \caption{(a,b) RIXS intensity maps and (c) RIXS loss spectra of the parent LaNiO$_2$ and NdNiO$_2$ compounds, together with the corresponding XAS spectra at the Ni $L_3$-edge [superimposed black curves in (a) and (b)] (from \cite{hepting20}). The data shows a main peak A and a lower energy shoulder A'. Both signatures are clearly seen in LaNiO$_2$, while A' is resolved only the RIXS loss spectra for NdNiO$_2$ as indicated by the arrow in (c). These features confirm the multiband character of these system that results from the from the hybridization between Ni-3$d_{x^2-y^2}$ states and the 5$d$ states of the rare earth. 
    }
    \label{fig:XAS}
\end{figure}

This question has been further addressed by X-ray spectroscopic techniques \cite{hepting20,lee20XAS}. The infinite-layer nickelates feature a main absorption peak similar to the cuprates (the A peak in Fig. \ref{fig:XAS}). This feature can be naturally ascribed to the presence of Ni-3$d$ states near the Fermi energy (see below). 
However, the spectrum contains an additional lower-energy feature (A' peak in Fig. \ref{fig:XAS}). 
This reveals the presence of additional states, and hence gives further evidence of the multiband character of infinite-layer nickelates. 
On the other hand, the lack of a pre-peak in the O $K$-edge suggests that the mixing between oxygen and nickel states is substantially weaker than in cuprates. This is in tune with the absence of a clear Mott or charge-transfer insulating behavior observed in the transport \cite{hwang20-EEELS}. In contrast to the Hall data, however, the changes in the X-ray spectra observed as a function of Sr-doping have been in turn interpreted in terms of doped holes residing in the Ni-3$d_{x^2-y^2}$ orbitals without necessarily invoking multiband effects \cite{lee20XAS}.

The single particle tunneling spectrum has been measured on (Nd,Sr)NiO$_2$ thin films deposited using MBE \cite{wen20a}. The spectrum was found to be inhomogenous across the sample, with different features at different locations of the sample. One of the predominant features correspond to a V-shape spectrum, which can be fitted to a $d$-wave gap with amplitude $\sim$ 3.9 meV. 
However, features corresponding to a full $s$-wave gap of about 2.35 meV are equally observed and, in some cases, even a mixture of the two. This surprising finding is now open to clarification which requires, in particular, a better understanding of the surface-specific local properties.

\section{Theoretical considerations}

Resolving the degree of connection between nickelate and cuprate superconductivity has been the one of the main motors of a rather frenetic theoretical activity. 
A number theoretical tools and ideas were already on the table for constructive confrontation. 
The electronic structure, in particular, has been revisited from different perspectives and for different purposes. 

\subsection{Single-particle picture}

\begin{figure}[t!]
    \includegraphics[height=.21\textwidth,valign=t]{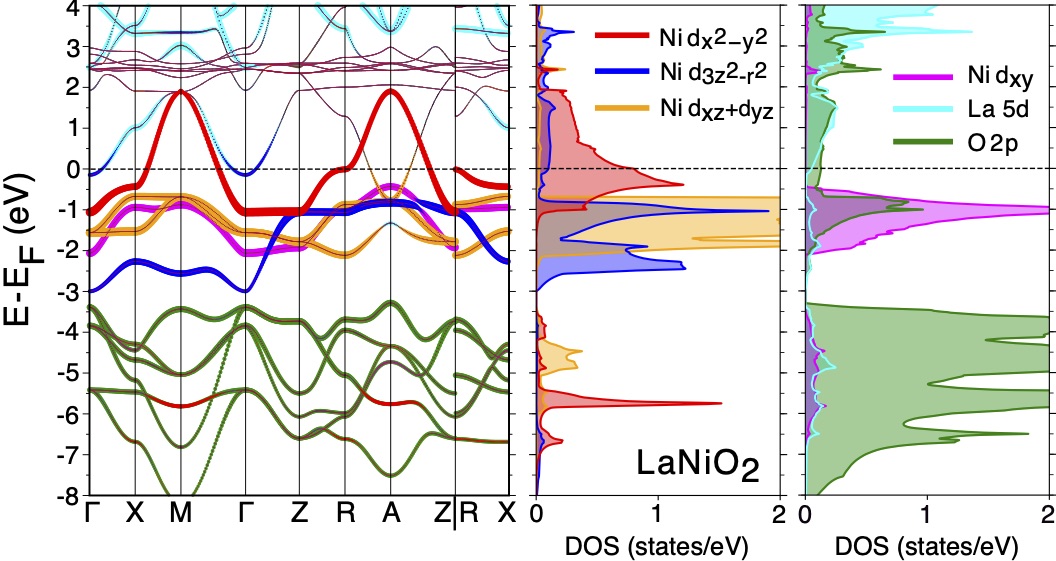}
    \includegraphics[height=.1825\textwidth,valign=t]{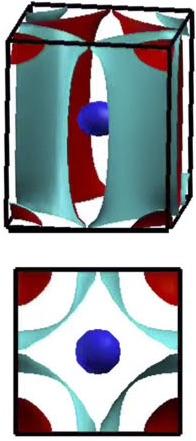}
    \caption{
    Electronic structure of the reference material LaNiO$_2$ obtained at the DFT level [fatband plot, density of states (DOS) and Fermi surface from \cite{botana20prx} and \cite{kotliar20prb}]. 
    In addition to the main Ni-3$d_{x^2-y^2}$ cuprate-like band crossing the Fermi level, the Ni-3$d_{z^2}$ states are not completely occupied in these systems. 
    Further, the system has a multiband character due to La-5$d$ states that ``self-dope'' the main Ni-3$d_{x^2-y^2}$ hole Fermi surface and give rise to additional electron pockets. The O-$2p$ sates, in turn, are comparatively far below the Fermi level.  
    }
    \label{fig:bands}
\end{figure}

At the single-particle DFT level, the calculations support the multiband picture of the infinite-layer nickelates in agreement with experimental data \cite{pickett-prb04,botana20prx,sakakibara20prl,hepting20,thomale20a,arita19,weng20topo,zhong20-prbDFT,jia20-DFT,cano20a,botana20magnetism,botana20doping, botana20prx}. 
These calculations reveal a large Fermi-surface sheet due to Ni-3$d_{x^2-y^2}$ holes akin to that in cuprates (see Fig. \ref{fig:bands}). 
However, the bandwidth of the corresponding band is comparatively reduced and its $c$-axis dispersion is much larger making this system  non-truly two-dimensional  \cite{pickett-prb04,cano20a}. Therefore, the analogy with cuprates has limitations already at this point. 
Furthermore, the Fermi surface displays additional electron pockets due to 5$d$ states associated to the rare-earth 5$d$ states (5$d_{z^2}$ at $\Gamma$ and 5$d_{xy}$ at A). 
This can be seen as a self-doping effect promoted by the hybridization of these formally empty states with the Ni-3$d$ bands. This effect is totally absent in the cuprates. 
In addition, the Ni-3$d_{z^2}$ states turn out to be partially occupied and additionally hybridized with the R-5$d$ ones. 
Thus, the full $e_g$-like=$\{d_{x^2-y^2},d_{z^2}\}$ sector of the Ni-3$d$ states becomes `active' in infinite-layer nickelates \cite{pickett-prb04,botana20prx}. 

The presence of $R$-5$d$ electrons is indeed compatible with the XAS spectrum as well as with the negative Hall coefficient measured experimentally in the parent compounds. 
Quantitative agreement with the Hall data, however, requires to somehow gap out the contribution from the main hole Fermi surface for the parent compounds \cite{botana20prx}. Similarly, the Uemura-plot phenomenology relating the superconducting $T_c$ to the magnetic penetration depth in the cuprates can be recovered in the nickelates only if the contribution from the main Fermi surface is somehow gaped out \cite{cano20a}. Note that the former refers to the transport in the normal state, while the latter does to superconductivity in its thermodynamic sense. Thus, these quantities provide complementary insights about the charge-carrier properties.

Beyond that, one important parameter that can be obtained from DFT calculations is the difference in charge-transfer energies. 
That is, the difference in on-site energies promoting charge from O-2$p$ to Ni-3$d$ orbitals. 
This difference is to be compared to the Coulomb-interaction energy-scale $U$ on the Ni sites, as this determines whether the system could be considered as charge-transfer or Mott-Hubbard-like according to the Zaanen-Sawatzky-Allen scheme \cite{zsa}. In spite of some discrepancies in absolute values among different works, there seems to be a consensus in that the charge transfer energy in these nickelates is larger than in cuprates 
 ($\sim$ 4~eV vs 2~eV), thereby precluding nickelate and cuprate from falling under the same category \cite{sawatzky20a}. 
The reduced bandwidth of the Ni-3$d_{x^2-y^2}$ states can also be understood from this increase in charge-transfer energies. These additional departures from the cuprate picture are again consistent with the XAS spectrum and the weakly insulating, yet metallic behavior observed in $R$NiO$_2$. 

These insights, essentially obtained from the comparison between experiments and single-particle DFT calculations, have set the stage and motivated a detailed analysis of the nickelate-specific electronic correlations.
In this respect, a methodological note is due on the DFT-based approaches. Undoubtedly, they provide important fundamental insights.
However, the state of the art is such that the $4f$-electrons introduced by the magnetic rare-earth elements cannot be treated on the same footing as the rest of states. 
This circumstance leaves LaNiO$_2$ as reference compound, in the sense that only here all power of a first-principles approach can be truly exploited. 
Otherwise, the $4f$-electrons are treated as core electrons ---so that Nd or Pr become electronically equivalent to La--- or LDA+$U$-corrected, which may not be totally justified.

\subsection{Many-body correlation effects}

\subsubsection{Cooper pairing}

The standard electron-phonon mechanism has been ruled out as the main reason for superconductivity in infinite-layer nickelates \cite{arita19}. 
Instead, repulsive interactions mediated by spin-fluctuations were right away argued to drive the Cooper pairing in these systems \cite{sakakibara20prl,thomale20a, wang20,dasgupta20prb}. 
Specifically, $d$-wave superconductivity was concluded from complementary random phase approximation (RPA) and fluctuation exchange (FLEX) calculations for many-body multi-orbital Hamiltonians in which the non-interacting part maps the relevant DFT bands. 
As for the interaction part, 
it is now customary to consider two-particle on-site interactions according to the Hubbard-Hund model Hamiltonian
\begin{align}
H_{int} &= 
U \sum_\alpha n_{\alpha \uparrow} n_{\alpha \downarrow} + U'\sum_{\alpha < \alpha'}  n_{\alpha }n_{\alpha' }  
\nonumber \\ &\quad 
+ J\sum_{\alpha < \alpha', \sigma , \sigma'}  
c_{\alpha \sigma}^\dagger c_{\alpha' \sigma'}^\dagger c_{\alpha \sigma'}c_{\alpha' \sigma} 
\nonumber \\ &\quad 
+J'\sum_{\alpha \not = \alpha'}  
c_{\alpha \uparrow}^\dagger c_{\alpha \downarrow}^\dagger 
c_{\alpha' \downarrow} c_{\alpha' \uparrow}, 
\label{Hint}
\end{align}
where $n_{\alpha \sigma} = c_{\alpha \sigma}^\dagger c_{\alpha \sigma}$ and $n_\alpha = n_{\alpha \uparrow} + n_{\alpha \downarrow}$ ($c_{\alpha \sigma}^\dagger$ creates electrons in orbital $\alpha$ with spin $\sigma$). Here $U$ and $U'$ represent intra- and inter-orbital Hubbard interactions, $J$ the Hund's rule exchange, and $J'$ the `pair hopping interaction'. In practice, these parameters are taken such that $U'=U-2J$ and $J'=J$ which corresponds to a spin rotational invariance for the overall two-body local interaction. 
These interactions were considered at both Ni and rare-earth sites in \cite{sakakibara20prl}, while the problem was simplified in \cite{thomale20a} by restricting Ni interactions to the intra-orbital Hubbard $U$ in accordance with their 3-band description. 
The results obtained from these two models are nevertheless compatible (see also \cite{wang20,dasgupta20prb}). In fact, in this scenario, the reduced $d_{x^2-y^2}$ bandwidth combined with its presumably large intra-orbital interaction $U_{x^2-y^2}$ may explain the reduced $T_c$ compared to cuprates \cite{sakakibara20prl}. 
This situation can be traced back to the aforementioned increase in charge-transfer energy. 
More recently, these results have been confirmed using advanced techniques in which the starting vertex is non-perturbative so that the local correlations are fully included \cite{held20-superconductivity}.

The same conclusion about the $d$-wave symmetry of the superconducting gap was reached in \cite{thomale20a} from a standard $t$-$J$ model constructed in a similar way for the Ni-3$d_{x^2-y^2}$ states. Further, the specific self-doping features of the nickelates have inspired an extended $t$-$J$ model that generically addresses the strong-coupling limit of similar multiband systems  \cite{vishwanath20prr}. This model is found compatible with Fermi-liquid behavior and $d$-wave superconductivity.
Alternatively, if the Hund's coupling $J$ between $d_{x^2-y^2}$ and $d_{z^2}$ Ni orbitals plays a dominant role, it has been pointed out that the Cooper pairing can be interpreted within a spin-freezing scenario as due to local-moment fluctuations, rather than to pure antiferromagnetic fluctuations \cite{werner20Hund}. 

\subsubsection{Electronic structure}

As mentioned above, infinite-layer nickelates in their normal state systematically show a weakly-insulating or bad-metal behavior, also in the overdoped regime (see Fig. \ref{fig:phase-diagram}). 
This is in clear deviation from canonical Mott physics and the charge-transfer-insulator to Fermi-liquid crossover that defines the high-$T_c$ cuprates.  
Yet, a modified correlated picture incorporating key multiband aspects is expected to provide important insights.

If electronic correlations remain relatively weak, then it is natural to revise the DFT picture by means of many-body perturbation theory in the first place. This enables the {\it ab initio} treatment of these correlations, which has been performed at the $GW$ level \cite{cano20b,werner20GWDMFT}. 
The resulting many-body picture thus includes the effect of the dynamically screened, but otherwise long-range Coulomb interaction. 
The low-energy physics remains essentially unaffected, with only small changes obtained in the interacting Fermi surface and in the quasiparticle spectral weights near the Fermi level. 
The Ni-3$d_{x^2-y^2}$ bandwidth reduces slightly while the Ni-3$d_{z^2}$ one increases as the O-2$p$ states are further shifted to lower energies ($1.5$~eV further down from the Fermi level) \cite{cano20b}. 
The latter, however, represents a rather substantial change, and hence suggests that the canonical charge-transfer-insulator picture is even more unlikely for the infinite-layer nickelates in the $GW$ framework. 
These changes are also tied to an important shift of the empty 4$f$ states \cite{cano20b}, which should be taken as a warning regarding their role in the overall physics of these materials.    

DFT+DMFT (dynamical mean-field theory) calculations according to the model interaction Hamiltonian \eqref{Hint} have been performed to further scrutinize the correlated nature of the different orbitals and clarify the multiband nature in $R$NiO$_2$ \cite{han20prbDMFT,werner20Hund,lechermann20prb,held20-superconductivity,millis20a,savrasov20a,lechermann20prx,savrasov20b,savrasov20c,kotliar20prb,kotliar20b}. 
In addition, DMFT has also been applied in combination with the quasiparticle self-consistent $GW$ approximation in a parameter-free fashion \cite{kotliar20a,werner20GWDMFT}.  
When it comes to the low-energy physics, the overall multiband picture remains robust and the results confirm the above trends. 
However, the effective mass renormalization or inverse quasiparticle weight $m^*/m = 1/Z$ undergoes substantial orbital-selective changes \cite{lechermann20prb,savrasov20a,kotliar20prb}. The Ni-$d_{x^2-y^2}$ band is found to have a tendency towards localization such that a Mott gap can eventually open if the Hubbard interaction is large enough  \cite{lechermann20prb, millis20a}. 
The $R$-5$d$ self-doping bands, in contrast, remain much more weakly correlated \cite{lechermann20prb, millis20a}. 

By modeling the system as a self-doped Mott insulator, a strong Kondo coupling between itinerant and localized carriers has been attributed to the low-temprature upturn observed in the resistivity data \cite{zhang20prbkondo}. 
The proposed model, however, tacitly assumes that such a localization is completely cuprate-like, while the actual picture is likely more complex given that the charge-transfer energy is much larger. 
In addition, the Ni-$d_{z^2}$ states are systematically found partially occupied and hence susceptible of charge fluctuations \cite{lechermann20prb,millis20a,kotliar20prb},  and may even cross the Fermi level under doping thereby supplementing the system with flatband features \cite{lechermann20}. This $d_{z^2}$ activity promotes the importance of a high-spin $d^{8}$ configuration with no $d^{10}$ involvement, in contrast to cuprates  \cite{lechermann20prb,millis20a,kotliar20prb}. Interestingly, the Ni-3$d$ total occupancy is found to be rather insensitive to hole-doping, as the changes in carrier density are absorbed as changes in the hybridization with $R$-5$d$ and O-2$p$ states \cite{savrasov20a,kotliar20prb,werner20GWDMFT, ku20doped}. 
Experimental XAS and RIXS data, however, has been interpreted slightly differently in terms of 
a single Ni-3$d_{x^2-y^2}$ orbital where the doped holes tend to reside in a low-spin configuration \cite{lee20XAS}. 

Thus, from the point of view of correlations, one current picture describes infinite-layer nickelates in terms of bad-metallic Ni-$e_g$ states coupled to itinerant $R$-5$d$ bands. 
Rather than to canonical Mottness, 
the bad metallic behavior here is ascribed to Hundness as the formally occupied Ni-3$d_{z^2}$ orbitals are found to participate in the low-energy physics in a Hund-assisted manner \cite{werner20GWDMFT, kotliar20prb}. An alternative picture describes these correlations as dominated instead by
the Ni-3$d_{x^2-y^2}$ orbital, thus suggesting that a one band (plus charge reservoir) Mott-Hubbard-like description of the low-energy physics may
be more appropriate \cite{karp2020_b}. 
This idea builds on the fact that the self energy and the spectral function display the characteristic forms expected in a Mott-Hubbard system, including the characteristic three-peak structure in the spectral function and particle-hole symmetric features in the self-energy above and below the Fermi energy. 
This defines an apparent Hubbard vs. Hund dichotomy in which the multiband aspects of infinite-layer nickelates are emphasized differently \cite{lechermann20}.

\subsection{Magnetism}

As mentioned above, no sign of long-range magnetic order has been reported so far for any $R$NiO$_2$ parent material \cite{mei19-XPSspinfluctuations, crespin83, hayward99, hayward2003, ikeda16}, although a recent NMR study shows the presence of antiferromagnetic fluctuations and quasi-static antiferromagnetic order in Nd$_{0.85}$Sr$_{0.15}$NiO$_2$ \cite{yu20}.
In fact, a magnetic ground state is consistently obtained in theoretical studies \cite{pickett-prb04, botana20prx,hepting20,mei20prr,chen20cm,yang20npjQM,savrasov20b,savrasov20c,lechermann20, pardo_afm}. 
Both spin-polarized DFT and DFT+DMFT calculations reveal a near degeneracy of various types of spin orders though, implying frustration of magnetic correlations in infinite-layer nickelates. 
In particular, using DFT+DMFT, first- and second-nearest neighbor exchange constants have been derived that put these materials in the regime of magnetic frustration upon doping within
a $J_1$-$J_2$ spin model \cite{savrasov20b}. 
This unanticipated frustration, that allows the suppression of magnetic order, arises due to the involvement of both 3$d_{x^2-y^2}$ and 3$d_{z^2}$ Ni orbitals. As such, this picture highlights the importance of in-plane spin fluctuations in understanding the physics of these materials and reinforce the requirement of an effective Ni-$d_{x^2-y^2}$ and $d_{z^2}$ two–band model to describe magnetic excitations in hole-doped $R$NiO$_2$  \cite{savrasov20c}.

\begin{figure*}[t!]
    \includegraphics[width=.85\textwidth]{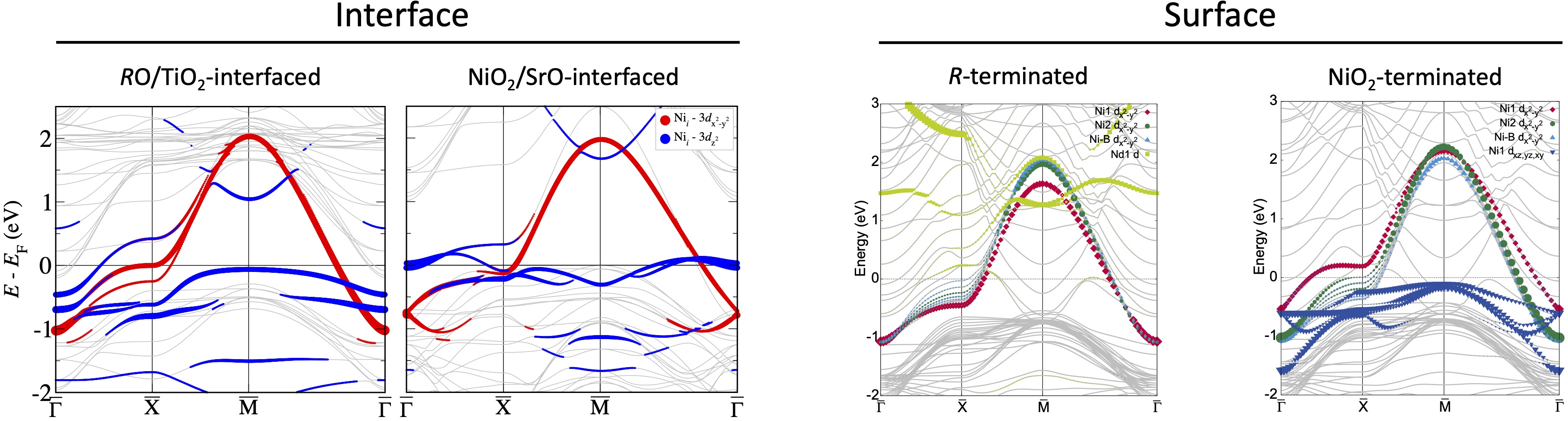}
    \caption{
    Electronic band structure of the infinite-layer nickelates at the $R$NiO$_2$/SrTiO$_3$ interface (left) and at the surface (right) for different atomic configurations (the colors highlight the main contributions of the interfacial/surface Ni-3$d$ states near the Fermi level; adapted from \cite{cano20c} and \cite{thomale20b}). 
    In both cases the nature of the self-doping effect ---obtained from the $R$-5$d$ states in the bulk--- changes or even disappear depending on the local atomic configuration. At the same time, other Ni-3$d$ states are pushed closer to the Fermi level and locally supplement the system with flatband features.}
    \label{fig:local-bands}
\end{figure*}

A complementary point of view is actually obtained \cite{choi20_2} via DFT calculations as the static antiferromagnetic ground state is characterized by the involvement of both $d_{x^2-y^2}$ and $d_{z^2}$ Ni bands. This state is peculiar in that it portrays a flat band,
one-dimensional-like van Hove singularity of $d_{z^2}$ character pinned at the Fermi level. 
This singularity makes the antiferromagnetic phase unstable
to spin-density disproportionation, breathing and half-breathing lattice distortions, and charge-density disproportionation. These flat-band instabilities should inhibit but not eliminate incipient antiferromagnetic tendencies at low temperature. 
A different point of view suggests that 5$d$ conduction electrons could screen the Ni spins, suppressing magnetism and giving rise to a Kondo effect like that seen in heavy fermion materials \cite{zhang20prbkondo}.

Beyond that, 
an intrinsic difference that could be behind differing properties across the $R$NiO$_2$ series, notably the emergence of superconductivity itself, 
is the presence/absence of magnetic rare-earth elements \cite{choi20_1,botana20magnetism}. 
All in all, magnetism in infinite-layer nickelates is still under intense study and further experiments will be 
crucial to shed some light on this problem. 

\subsection{Interfacial and surface effects
}

To date, superconducting samples of infinite-layer nickelates have only been obtained thanks to thin-film growth techniques. The surface/interfaces of these epitaxial thin-films are readily available for experimental characterization. Thus, the question of whether their electronic properties match the bulk picture ---so that superconductivity can be preserved and/or supplemented with additional local features \cite{wen20a}--- becomes fully pertinent. 

This question was first addressed by means of DFT calculations for $R$NiO$_2$/SrTiO$_3$ superlattices in \cite{cano20c}, and shortly after in \cite{pentcheva20a}. 
The epitaxial growth of these heterostructures can in principle yield different atomic boundaries between the sample and the substrate. Consequently, the first important question to clarify is which interface is actually realized in these systems \cite{cano20c}. 
The most obvious configuration corresponds to a fully reduced nickelate having its ($R$O~$\to)$~$R$ layer directly on top of the TiO$_2$-terminated substrate. 
This configuration, however, was found to be energetically unstable. 
The reason is that, even if the reduction process is effective in the bulk, the removal of the interfacial apical oxygen turns out to be much harder. This is just a local-scale manifestation of the thermodynamic fragility of these phases. 
Thus, the infinite-layer nickelate prefers to face a $R$O layer to the substrate, and the same conclusion holds even for a direct-growth process. 
This prediction is totally in tune with subsequent STEM images in which these interfacial oxygens are visible (see e.g. \cite{hwang20aspects}). 
The issue, however, is not completely resolved since, depending on the growth and topotactic-reduction conditions, NiO$_2$/Sr and NiO$_2$/SrO interfaces may also be obtained as metastable configurations. Also, similarly to the topotactic H \cite{held20topotaticH}, the oxygen binding energy can be different for different rare-earth atoms, which may render these configurations even more viable.   

In any case, this sort of `chemical' reconstruction
was further shown to produce drastic changes in the electronic structure at the interface \cite{cano20c}. 
In fact, the interfacial chemical reconstruction according to the $R$ $\to$ Sr $\to$ $R$O $\to$ SrO sequence is to some extent equivalent to localized hole doping. 
This local doping was found to deplete the self-doping $R$-5$d$ states at the interface. Specifically, the $R$-5$d$ states are first replaced by Ti-3$d$ ones, which are then pushed above the Fermi level for the SrO configuration (see Fig. \ref{fig:local-bands}). 
At the same time, the intefacial Ni-3$d_{z^2}$ states are driven closer to the Fermi energy so that they manifestly participate in the low-energy physics. 
Thus, the Kondo-lattice features are expected to be fundamentally different at the interfaces where, in addition, the Ni-$e_g$ sector will likely be fully active. 
Besides, this sector is supplemented by a markedly flatband character of the interfacial Ni-3$d_{z^2}$ states, so that interface-specific correlation effects may be promoted \cite{cano20c}. 

This picture was subsequently confirmed for thin films with asymmetric boundaries \cite{pentcheva20a,dagotto20,zhong20prbinterfaces}. 
In that case, the different polar discontinuities yield an effective built-in electric field across the film. The screening of this field is then an additional ingredient determining the eventual atomic and electronic reconstruction. 
Thus, polar layers can be formed at the surface and at the interface \cite{dagotto20}. These layers show antiparallel NiO$_2$ displacements, but otherwise are decoupled. At the interface with the substrate, the calculations reveal the formation of a two-dimensional electron gas extending over several layers together with the aforementioned depletion of the self-doping $R$-5$d$ states  \cite{dagotto20,pentcheva20a}. In addition, the combined effect of magnetism ($G$-AFM order) and correlations has been considered at the DFT+$U$ level \cite{dagotto20}. 
This combination has been found to enhance the itineracy of the 
Ni-3$d_{z^2}$ orbitals at the interface with the substrate, while the magnetism is essentially suppressed at the surface to vacuum.
The focus has been put on the surface properties in \cite{thomale20b}. Specifically, the effect of the nickelate termination anticipated in \cite{cano20c} is considered in detail. Thus, it is confirmed that different terminations yield different electronic structures also at the surface (see Fig. \ref{fig:local-bands}). This is further shown to modify qualitatively the corresponding Fermi surface. As a result of this modification, it is argued that the $d$-wave superconducting gap expected for the bulk may transform into a $s_\pm$-wave one at the NiO$_2$-terminated surface. This provides a rather natural explanation to the local changes in the tunneling spectrum observed in \cite{wen20a}. Further, it suggests that a surface $s+id$-wave state may also be realized under the appropriate conditions. 

\section{Conclusions and perspectives}

The recent discovery of superconductivity in infinite-layer nickelates has created intense excitement. 
These systems have been rapidly scrutinized from many different angles, using a battery of experimental and theoretical tools. 
The initial motivation of drawing analogies with the high-$T_c$ cuprates has thus been surpassed. Instead, the accumulated results have now consolidated these systems as a new class of unconventional superconducting materials.   

Specifically, the rare-earth infinite-layer nickelates have been confirmed to host a distinct multiband interplay, on top of which electronic correlations build and determine the main properties of these systems. 
This interplay is present already within the Ni sector itself, as not only the Ni-3$d_{x^2-y^2}$ states but also the Ni-3$d_{z^2}$ ones are found to be active. 
This further introduces specific correlation effects and the bad metallic, or weakly insulating behavior is now understood as a direct manifestation of these correlations. However, a Hubbard vs Hund dichotomy has emerged that is yet to be clarified.   
In addition, the rare-earth states introduce extra specific ingredients such as the self-doping effect and a 4$f$-ness that may qualitatively be even more important. 
When it comes to the central question, that is, the emergence of superconductivity in these materials, it has been ascribed to spin fluctuations (in a broad sense), and there is now experimental evidence of incipient antiferromagnetic order.      

These are now well-established traits of the infinite-later nickelates. 
Additional progress to further clarify these aspects, as well as the actual superconducting properties beyond $T_c$ can be naturally expected \cite{hwang21nphys}.
At the same time, there are many other crucial questions yet to be answered. Concerning issues related to the dependence of superconductivity on the rare-earth element, much more could be learned if additional infinite-layer variants could be grown and doped. In this context, it is fundamental to determine whether lack of superconductivity in the LaNiO$_2$
reference material is intrinsic or not. 
Another crucial question is why thin films are superconducting while bulk samples are not.
This may simply be due to sample quality and doping control. However, there is also a $\sim$3\% $c$-axis lattice constant difference between bulk and superconducting thin films. 
If that difference alone explains superconductivity, that could provide additional evidence of the explicit role of the Ni-3$d_z^2$ orbital (the same argument would effectively apply to LaNiO$_2$ in comparison to $R$NiO$_2$ with smaller $R$ ions). 

More importantly, 
a crucial issue to address is: 
are the rare-earth infinite-layer nickelates a solitary beast or just the tip of the iceberg? 
That is, is there a whole new family of nickel-based unconventional superconductors waiting to be discovered? 
This question currently motivates the experimental and computational search of new alternative materials \cite{botanaprm, zhang_nat_phys, poltavets1, poltavets2, Poltavets_3, Poltavets_4,lorenzana19,arita20prb,cano20d, karp2}.
As mentioned above, $R$NiO$_2$ are the $n=\infty $ member of the larger series $R_{n+1}$Ni$_n$O$_{2n+2}$. 
Other members of this layered nickelate family are obtained via a similar oxygen reduction from the corresponding Ruddlesden-Popper phases \cite{Greenblatt_RP1, GreenblattRP2}. The $n=2$ and $3$ materials, in particular, have been known for a while and, similarly to the $n=\infty $ ones, have also been discussed as candidate superconductors 
\cite{botanaprm, zhang_nat_phys, poltavets1, poltavets2, Poltavets_3, Poltavets_4}. 
Recently, La-based $n=4$-$6$ parent Ruddlesden-Popper phases have also been synthesized. Reduction of these compounds is particularly promising as they would realize $d$-electron counts that can be directly mapped into the dome area of filling. In addition, current epitaxial growth techniques can be exploited as an alternative route to engineer Ni-based heterostructures mimicking the $R_{n+1}$Ni$_n$O$_{2n+2}$ series with the advantage of better sample quality and doping control \cite{cano20c}. The infinite-layer case itself has proven to be a challenging but successful example in this respect.  

Shedding light on these issues will not only help understanding
superconductivity in these 
specific low-valence layered nickelates but will also provide new perspectives about the nature of unconventional superconductivity in general. 

\noindent{\it Acknowledgments.---} 
We dedicate this review on this 1-year-old topic to I. E. Dzyaloshinskii in celebration of his 90th birthday. 
We thank M. R. Norman and X. Blase for useful comments. A. B. acknowledges the support from NSF DMR 2045826. 

\bibliography{bib.bib}

\end{document}